\documentclass[12pt]{article}
\usepackage{amscd,amsmath,amssymb}
\usepackage[cp866]{inputenc}
\usepackage[dvips]{graphicx}
\title{Dynamic general covariance of physical systems}
\author{Sergey S. Kokarev\thanks{logos-center@mail.ru}}
\date{ RSEC Logos (Yaroslavl)}

\begin{document}

\maketitle

\begin{abstract}
One unusual property of dynamic
systems, whose state is characterized by a set of scalar
dynamic variables satisfying a system of differential equations of a general form, is considered.
 This property is related to the behavior of equations (optionally covariant)
with respect to coordinate diffeomorphisms: the equations, in a sense, retain their form on their solutions.
More precisely, non-covariant addends to the equations of such systems
always exactly reduced in any order of  perturbation theory by
solutions of unperturbed (initial) equations. This property
demonstrated by a set of simple illustrative examples.
Various aspects of the dynamic covariance are discussed.
\end{abstract}

\section{Introduction}

General Relativity (GR) and tensor formalism have introduced into modern physics
many bright and important new ideas \cite{kok1}. One of them is  {\it
 general covariance of fundamental physical laws.}
Since coordinates in GR
do not have the physical meaning of observed or measured quantities (they are only more or less convenient event labels
 that we are free to choose to a large extent arbitrarily) the requirement of covariance is a necessary condition for objectivity
of physical laws. Let
\begin{equation}\label{EQ}
\hat{\mathcal{E}}_x(\Phi_x)=0
\end{equation}
be the equation, describing a physical system, where $\Phi_x$ is the set of its dynamical variables,  $\mathcal{E}_x$ ---
physical operator, whose kernel is the solution to (\ref{EQ}), $x$ is coordinate system, to which the physical  system is related.
Mathematically covariance of (\ref{EQ})
is expressed by the following requirements: under an arbitrary
diffeomorphism
\begin{equation}\label{chcoord}
x\to x'=f(x),
\end{equation}
where $f$ is the set of smooth functions with non-zero jacobian and smooth inversion,
the equation (\ref{EQ}) is equivalent to
\begin{equation}\label{EQ1}
\hat{\mathcal{E}}_{x'}(\Phi'_{x'})=0,
\end{equation}
where $\Phi'_{x'}$ are transformed by the rules of tensor algebra to the new coordinate system (hatch over $\Phi$ stresses
that dynamical variables can change its kind and components under coordinate transformation), while the operator
$\hat{\mathcal{E}}_{x'}$ conserves its form (hatch over $\hat{\mathcal{E}}_{x'}$ is absent).

If the covariance condition of (\ref{EQ})
is satisfied  not for an arbitrary  coordinate transformations (\ref{chcoord}), but
for some of its subgroup, it is said that the equation (\ref{EQ})
covariant with respect to this subgroup (or is specially covariant).  General discussion of the covariance of physical equations and its
roles in physical theories can be found in \cite{giulini}.
In geometry and physics
there are many examples of equations with general or special covariance.
For example, the equation:
\begin{equation}\label{EQ2}
\partial_\mu\Phi=0
\end{equation}
is general covariant, if $\Phi$ is general covariant scalar, and it is only affine-covariant, if $\Phi$ is tensor. The generally covariant generalization
of  (\ref{EQ2}) is the equation
\begin{equation}\label{EQ3}
\nabla\Phi=0,
\end{equation}
where $\nabla$ is covariant derivative, associated with some connection over the manifold, which describes dynamics of the system. Wave equation at the form:
\begin{equation}\label{EQ4}
(\partial_t^2-\overrightarrow{\partial}_x^2)\Phi=0
\end{equation}
even in case of scalar field will be covariant only with respect to  conformal group of Minkowski space-time.
 Generally covariant form of  (\ref{EQ4}) reads as follows:
\begin{equation}\label{EQ5}
(\nabla\cdot\nabla)\Phi\equiv\frac{1}{\sqrt{|g|}}\partial_\alpha(\sqrt{|g|}\bar g^{\alpha\beta}\partial_\beta\Phi)=0,
\end{equation}
where $g$ is metric over the manifold, $\bar g$ is inverse metric, $|g|=|\det\, g|.$

The coordinate transformations (\ref{chcoord}) have another
interpretation which is inspired by analogies from
physics of continuous media. Interpretation of formulas (\ref{chcoord}) as labels shifts
 implies that for an arbitrary point $p$ which had a label $x,$ new
label will be $ x'= f (x), $ so that the point $ p $ itself remains immobile (passive
viewpoint).
In continuous media physics transformations of the type (\ref{chcoord})
describe the deformation of a continuous media when coordinate labels are attached to its particles.
In this case, we assume that the particle of the medium labeled $x$
after deformation takes the position $x '= f (x)$  (active point of view or Lagrange picture).

In present paper we are going to investigate one curious property of dynamic
systems whose state is characterized by a set of scalar
dynamic variables. Of course, as the considered simple
examples show, an arbitrary equation or a system of equations of the form
(\ref{EQ}) does not have the property of general covariance in the sense of the formulas
(\ref{EQ})-(\ref{EQ1}). It turns out, however, that such systems
possess the property of {\it dynamic general covariance}: they retain their form on solutions to original equations.
In other words, non-covariant addends to the equations of such systems
always exactly reduced in any order of perturbation theory by
solutions of unperturbed (initial) equations. First we illustrate this property
by a very simple linear first-order system (the Sec.
\ref{lin1}), then by the example of an oscillator with damping (the
Sec.
\ref{lin2}), then using the example of general dynamic system,
described by an ordinary differential equation of the $ n $-th
order with variable coefficients (the Sec. \ref{lin3}). The case of a nonlinear system
is considered in the Sec. \ref{nonl}, of the wave equation --- in
Sec.
\ref {wave}, of a second-order perturbation  --- in the Sec.  \ref{second}.
The Sec. (\ref{gen}) is devoted to general review, summarizing all
previously considered cases. In the Conclusion we discuss some features and possible
physical interpretations of dynamic covariance.

\section{Linear system of first order}\label{lin1}

We will start by studying the transformational properties of a very simple
equation:
\begin{equation}\label{E1}
\dot x(\tau)=f(\tau),
\end{equation}
where $x $ is a scalar dynamic quantity, $ \tau $ is (for
definiteness)
time parameter, $ f $ --- scalar function ("source" \,),
dot denotes differentiation by argument. Suppose that an experimenter who measures $ x (\tau), $
had a suspicion  that his clock is experiencing some variation in its rate compared to a more stable standard. Such a variation
can be described by dependence of the form:
\begin{equation}\label{var1}
\tau=t+\epsilon(t),
\end{equation}
where $ \tau $ is the time related to the clock of the standard, $ t $ is time measured by laboratory clock,
$ \epsilon (t) = \tau-t $ is the shift function that we assume
to be small: $ \epsilon (t) \ll t. $ The equation (\ref{E1}) was deduced in
assumption of a stable time parameter, so if
the experimenter uses a laboratory clock with a shift function
$ \epsilon (t), $ he needs to rewrite the equation in terms of
laboratory time $ t. $ Given the smallness of $ \epsilon $ necessary
substitutions have the form:
\begin{equation}\label{tr1}
x(\tau)=x(t)+\dot x(t)\epsilon(t)+o(\epsilon);\quad f(\tau)=f(t)+\dot
f(t)\epsilon+o(\epsilon);\quad
\end{equation}
\begin{equation}\label{trd}
\frac{d}{d\tau}=\frac{1}{1+\dot
\epsilon(t)}\frac{d}{dt}=(1-\dot\epsilon(t))\frac{d}{dt}+o(\epsilon).
\end{equation}
Substituting  (\ref{tr1})-(\ref{trd}) into (\ref{E1}), we obtain after rather simple transformations up to  $o(\epsilon)$:
\begin{equation}\label{E1a}
\dot x=f-\epsilon(\ddot x-\dot f).
\end{equation}
Of course, the equation (\ref{E1}) have changed its form\footnote{For its covariance the value $ f $ in the right-hand side
should be a 1-dimensional vector field on
the one-dimensional manifold $\mathbb{R},$ parameterized by the parameter
$ \tau $ --- see the discussion in the Conclusion.}: the additional term in the right-hand side of (\ref{E1a}) is proportional to $ \epsilon, $ and is
 similar to inertia forces
in mechanics that arise as addends to physical forces in
non-inertial reference frames. Here and below we will call such addends  {\it deformation perturbations}
(now we are talking about deformations of the laboratory
clocks rate). The smallness of $ \epsilon $ implies the application of perturbation theory
for  finding corrections to the dependence $ x_0 (\tau), $ obtained as
solution to the original equation (\ref{E1}). When $ \epsilon = 0 $ (\ref{E1a})
goes (as it should be)  into (\ref{E1}), so zero  approximation
of solutions to (\ref{E1a}) coincides with $ x_0 (t). $ Assuming further
 $ x_1 (t) = x_0 (t) + \delta (t) $ with the initial condition $ \delta (0) = 0 $ (the value of $ x (0) $
 is controlled by the experimenter, and the zeroes of the laboratory and reference
 clocks  can always be taken the same by definition), we obtain the equation for
 $\delta $:
\begin{equation}\label{corr1}
\dot\delta=\epsilon(\ddot x_0-\dot f).
\end{equation}
However, the right-hand side of the equation (\ref {corr1}) is identically zero over zero approximation to the solution due to the equations of motion
in zero approximation:
\begin{equation}\label{ident1}
\dot x_0- f=0\Rightarrow\ddot x_0-\dot f=\frac{d}{dt}(\dot x_0-\dot
f)=0.
\end{equation}
By  (\ref{ident1}) the equation  (\ref{corr1}) on  $\delta$ takes the form:
\begin{equation}\label{corr1a}
\dot \delta=0,
\end{equation}
that under $\delta(0)=0$ leads to  $\delta(t)=0.$
We conclude that {\it small perturbation of laboratory clock rates for   the  system (\ref {E1})
 does not manifest itself as small perturbations of the dependence $ x (t). $}
Mathematically this statement is equivalent to the covariance
property  for equation (\ref {E1}) in weakened {\it
dynamic}
sense: {\it the equation (\ref{E1}) retains its form and solutions in case of weak
deformations  of parameter $ \tau $
over solutions to the original (non-transformed) equation.}

Our conclusion seems somewhat paradoxical: it's obvious that {\it some process have to be described by different dependencies when
time counting system is deformed} --- this dictates by common sense and
simple explicit examples. We defer discussion of this paradox to
Conclusions. For now, let us turn to the following more complicated  examples.

\section{Oscillator with damping}\label{lin2}

Consider an oscillatory system, which itself can be used as
reference standard of time and frequency. Such a system in a commonly
used  notations  will be
described by a second-order differential equation of the form:
\begin{equation}\label{E2}
\ddot x(\tau)+2\gamma\dot x(\tau)+\omega_0^2x(\tau)=f(\tau),
\end{equation}
where $ x $ is the scalar oscillating dynamic variable, $ \gamma $
is damping coefficient, $ \omega_0 $ is proper frequency of the system, $ f $ is  generalized
external force.
Repeating the logic of reasoning of the previous section related to
formula (\ref{var1}), we come to the need to rewrite
equation (\ref{E2}) in terms of laboratory time $ t. $ Replenishing
for this purpose formulas (\ref{tr1}) by the law of transformation of the second
derivative:
\begin{equation}\label{trd2}
\frac{d^2}{d\tau^2}=\frac{1}{1+\dot
\epsilon(t)}\frac{d}{dt}\frac{1}{1+\dot \epsilon(t)}\frac{d}{dt}=\frac{d^2}{dt^2}-2\dot\epsilon\frac{d^2}{dt^2}-\ddot \epsilon\frac{d}{dt}+o(\epsilon).
\end{equation}
and substituting (\ref{tr1})-(\ref{trd}) and  (\ref{trd2}) into (\ref{E2}), after rather simple calculations we obtain up to
$o(\epsilon)$:
\begin{equation}\label{E2a}
\ddot x(t)+2\gamma\dot x(t)+\omega_0^2x(t)=f(t)-\epsilon(\dddot x(t)+2\gamma\ddot x(t)+\omega^2_0 x(t)-\dot
f(t)).
\end{equation}
The deformational perturbation of the equation of forced oscillations in the right-hand side of (\ref{E2a}) again
turned out to be proportional to the derivative of the equations of motion and
we come
again to the previous paradoxical conclusion,   that {\it small time rate perturbations of the laboratory clock does not lead to small
  perturbations of the law of oscillations.} In other words, the principle
  of dynamic covariance holds for oscillatory systems.

\section{Arbitrary linear dynamical system}\label{lin3}

To make sure that the detected effect is not
exceptional feature of simple systems that are described
by linear differential equations with constant coefficients of order not higher
second, consider a general system with an equation of the form:
\begin{equation}\label{E3}
\mathcal{D}^n_\tau(x)=0,
\end{equation}
where
\begin{equation}\label{DD}
\mathcal{D}^n_\tau(x)\equiv\sum\limits_{k=-1}^na_k(\tau)D^k_\tau (x(\tau)),
\end{equation}
\begin{equation}\label{D}
{D}^k_\tau(x)\equiv
\left\{
\begin{array}{lr}
\displaystyle \frac{d^kx}{d\tau^k},& k=1,\dots,n\\
x,& k=0\\
1, & k=-1,
\end{array}
\right.
\end{equation}
and $a_{-1}(\tau)=-f(\tau).$
Deformation of $ \tau $ by (\ref{var1}) entails deformation
of dynamic variable $ x (\tau) = x (t) + \delta x (t) $ and of the operator $ \mathcal{D}_{\tau}^n\to \mathcal{D}_t^n + \delta \mathcal{D}^n_t,$
that after substituting into the equation (\ref{E3}) leads to
deformed equation:
\begin{equation}\label{E3a}
\mathcal{D}^n_t(x(t))=-\mathcal{D}^n_t(\delta
x(t))-\delta\mathcal{D}^n_t(x(t)).
\end{equation}
According to  (\ref{tr1}) $\delta x(t)=\epsilon\dot x+o(\epsilon).$ Calculation
of  $ \delta\mathcal{D}^n_t $
is carried out in three steps. In first, by (\ref{trd}) and
recurrence ratio:
\[
D^{k}_\tau=D_\tau(D^{k-1}_\tau),
\]
we find:
\begin{equation}\label{Dtr1}
D^{k}_{\tau}(x(t))=D_t^{k}(x)+D_t(\delta D^{k-1}_t(x))-\dot\epsilon
D^{k}_t(x)
\Rightarrow
\end{equation}
\[
\delta D^{k}_{\tau}=D_t(\delta D^{k-1}_t)-\dot\epsilon
D^{k}_t+o(\epsilon).
\]
Then, sequentially revealing the resulting recurrence relation
(\ref{Dtr1}), we obtain  the general formula:
\begin{equation}\label{Dtr1a}
\delta D^{k}_{t}(x)=-\sum\limits_{s=1}^{k}D^{k-s}_t(\dot\epsilon
D^{s}(x)).
\end{equation}
With using Leibniz rule:
\begin{equation}\label{Leib}
D^k_t(x\cdot y)=\sum\limits_{p=0}^kC^k_pD^p_t(x)D^{k-p}_t(y)
\end{equation}
($C^k_p=k!/p!(k-p)!$ are binomial coefficients) and after changing the order of summation,
the (\ref{Dtr1a}) is reduced to the following final form\footnote{In calculations of (\ref{Dtr1b}) the identity
\[
\sum\limits_{s=1}^{k-p}C^{k-s}_p=C^k_{p+1},\quad k\ge p+1,
\]
has been used. It is easily proved by induction.}:
\begin{equation}\label{Dtr1b}
\delta
D^{k}_{t}(x)=-\sum\limits_{p=0}^{k-1}C^k_{p+1}D^{p+1}_t(\epsilon)
D^{k-p}_t(x).
\end{equation}
The formula  (\ref{Dtr1b}) is valid for any $k\ge1.$ By
(\ref{D}) $\delta D^0=0,$ $\delta D^{-1}=0.$

Lets now  take into account the perturbations of the coefficients $a_k(\tau)$ in (\ref{DD}):
\begin{equation}\label{ak}
\delta a_k(t)=
\left\{
\begin{array}{lr}
\epsilon(t)\dot a_k(t),& k\ge0;\\
-\epsilon(t)\dot f(t)& k=-1.
\end{array}
\right.
\end{equation}

Putting it all together, we get for the perturbed equation (\ref{E3a}):
\begin{equation}\label{E3b}
\mathcal{D}^n_t(x)=-\sum\limits_{k=0}^na_k D^k_t(\epsilon\dot
x)-\sum\limits_{k=-1}^n\epsilon \dot a_k D^k_t(x)+\sum\limits_{k=1}^na_k \sum\limits_{p=0}^{k-1}C^k_{p+1}D^{p+1}_t(\epsilon)
D^{k-p}_t(x)).
\end{equation}
Using in the first term of the right-hand side (\ref{E3b}) the formula
(\ref{Leib}) and highlighting separately terms proportional to
$ \epsilon $:
\[
\sum\limits_{k=0}^na_k D^k_t(\epsilon\dot
x)=\sum\limits_{k=0}^na_k \sum\limits_{p=0}^{k-1}C^k_{p+1}
D^{p+1}_t(\epsilon)
D^{k-p}_t(x)+\sum\limits_{k=0}^n\epsilon a_k D^k_t(\dot
x)
\]
after substitution into (\ref{E3b}) and reduction of similar terms we obtain:
\begin{equation}\label{E3c}
\mathcal{D}^n_t(x)=-\epsilon\frac{d}{dt}\left(\mathcal{D}^n_t(x)\right).
\end{equation}

We see that in the case of a general linear system, the infinitesimal deformations of the time parameter lead to deformation perturbations
 of the equation proportional to
the full derivative of the unperturbed equation. In other words, {\it dynamic covariance is preserved for general linear
systems} .

\section{Non-linear systems}\label{nonl}

Now lets check whether the property of dynamic covariance extends to
nonlinear systems. For this purpose\footnote{Non-linear equations of a higher order are reduced to a system of equations of the type (\ref{nl1}).}
consider the equation of the form:
\begin{equation}\label{nl1}
\dot x(\tau)=\Phi(\tau,x).
\end{equation}
Using  (\ref{trd}) and the decomposition:
\begin{equation}\label{trf}
\Phi(\tau, x(\tau))=\Phi(t,x)+\Phi_{,x}\epsilon\dot
x+\Phi_{,t}\epsilon,
\end{equation}
we immediately obtain:
\begin{equation}\label{nl1a}
\dot x(\tau)-\Phi(\tau,x)=-\epsilon\frac{d}{dt}(\dot x-\Phi)
\end{equation}
--- the equation, expressing dynamic covariance property for non-linear system.

\section{Partial derivatives equations}\label{wave}

In fact, the property of dynamic covariance characterizes
the more general class of equations than ordinary differential ones. As an example, consider a 2-dimensional wave equation:
\begin{equation}\label{wave1}
u_{,\tau\tau}-u_{,\chi\chi}=J(\tau,\chi),
\end{equation}
($J$ is source), where the both coordinates are deformed:
\begin{equation}\label{deftx}
\tau=t+\epsilon_1(t,x);\quad \chi=x+\epsilon_2(t,x).
\end{equation}
Inverse to (\ref{deftx}) relations up to $o(\epsilon)$
have the form:
\begin{equation}\label{deftxi}
t=\tau-\epsilon_1(\tau,\chi);\quad x=\chi-\epsilon_2(\tau,\chi).
\end{equation}
Motivated by considerations of an experiment with distorted rulers and
clocks, we transform the equation (\ref{wave1}) to the variables $ t, x $,
assuming that $ u $ and $ J $ are scalar quantities.
Using (\ref{deftx})---(\ref{deftxi}) after simple calculations we get:
\begin{equation}\label{ut}
u_{,\tau\tau}=u_{,tt}+\delta
u_{,tt}-\epsilon_{1,tt}u_{,t}-2\epsilon_{1,t}u_{,tt}-\epsilon_{2,tt}u_{,x}-2\epsilon_{2,t}u_{,tx};
\end{equation}
\begin{equation}\label{ux}
u_{,\chi\chi}=u_{,xx}+\delta
u_{,xx}-\epsilon_{2,xx}u_{,x}-2\epsilon_{2,x}u_{,xx}-\epsilon_{1,xx}u_{,t}-2\epsilon_{1,x}u_{,tx};
\end{equation}
\begin{equation}\label{dJ}
\delta u=u(t,x)+u_{,t}\epsilon_1+u_{,x}\epsilon_2;\quad \delta
J=J(t,x)+J_{,t}\epsilon_1+J_{,x}\epsilon_2.
\end{equation}
Substituting all that into (\ref{wave1}) and reducing similar terms, we go to perturbed equation of the kind:
\begin{equation}\label{wavedef}
u_{,tt}-u_{,xx}-J(t,x)=-\epsilon_1(u_{,tt}-u_{,xx}-J(t,x))_{,t}-\epsilon_2(u_{,tt}-u_{,xx}-J(t,x))_{,x},
\end{equation}
from which it follows that {\it dynamic covariance takes place also
in a field theory.}

%

\section{The second order}\label{second}

Finally, we verify that the property of dynamic covariance
is not a consequence of infinitesimal deformations of time and
space. For this purpose, we return to the simplest system from the
Sec. (\ref{lin1}) and take into account the perturbation of the equation (\ref{E1}) in the second
order. For this purpose, we use the formulas:
\begin{equation}\label{second1}
x(\tau)=x(t)+\epsilon\dot x(t)+\frac{1}{2}\epsilon^2\ddot x(t)+o(\epsilon^2);\quad
\frac{d}{d\tau}=\frac{1}{1+\dot\epsilon}\frac{d}{dt}=(1-\dot\epsilon+\dot\epsilon^2)\frac{d}{dt}+o(\epsilon^2.)
\end{equation}
Substituting (\ref{second1}) into (\ref{E1}), after rather simple
transformations, we obtain:
\begin{equation}\label{E1second}
\dot x(t)-f(t)=-\epsilon\frac{d}{dt}(\dot
x-f)-\frac{1}{2}\epsilon^2\frac{d^2}{dt^2}(\dot x-f).
\end{equation}
We see that the {\it deformation perturbations of the equation (\ref{E1}) vanish on the solutions of the unperturbed system
in the first and second orders independently.}
Thus, the {\it dynamic covariance of the equation (\ref{E1})
is preserved when accounting for the second order.}

In fact, one can verify by the similar direct calculations that in all the examples considered,
second-order accounting does not violate dynamic covariance
of equations in these examples.

\section{General consideration}\label{gen}

The considered examples suggest that the property
of dynamic covariance is a common one for any
dynamical systems described by ordinary differential systems
or partial derivatives equations in any order of perturbation theory. General reasoning
which detects this and generalizes the explicit calculations of all previous
examples  is as follows\footnote{For simplicity, we restrict ourselves to the case of systems described by a single partial differential equation.
Generalization to the case of systems is trivial.}. Let, as in (\ref{EQ})
\begin{equation}\label{GEQ}
\hat{\mathcal{E}}_x(\Phi_x)=0
\end{equation}
is partial differential equation with respect to
the coordinate set $x, $ $ \Phi $ is a scalar field.
Let's make a deformation (finite) of coordinates of the form:
\begin{equation}\label{GEQ1}
x=x'+\epsilon(x'),
\end{equation}
where $ \epsilon (x') $ is the deformation field. After substituting some
functions $ \Phi_x $ (optional solutions) into the operator $ \hat{\mathcal{E}}_ x $
in the left side (\ref{GEQ}), the expression $ \hat{\mathcal{E}}_ x (\Phi_x) $ can
regarded as some function $\mathcal {E} (x), $ which under
assuming of its analyticity,  can be expanded into a Taylor series in
neighborhood of the point $x'$:
\begin{equation}\label{GEQ2}
\mathcal{E}(x)=\sum\limits_{s=0}^{\infty}\partial^s\mathcal{E}(x')\frac{\epsilon^s(x')}{s!},
\end{equation}
where $ s $ is the collective summation index reflecting the general
structure of
Taylor series for a function of many variables. Denote now
$ \mathcal{E}_0 (x) \equiv \hat{\mathcal{E}}(\Phi_0), $ where $ \Phi_0$
is
solution to (\ref{GEQ}). Then instead of (\ref{GEQ2}) we have:
\begin{equation}\label{gen1}
\mathcal{E}_0(x)=\sum\limits_{s=0}^{\infty}\partial^s\mathcal{E}_0(x')\frac{\epsilon^s(x')}{s!}
\end{equation}
--- an expression generalizing the formulas (\ref{E1a}), (\ref{E2a}),
(\ref{E3c}), (\ref{nl1a}), (\ref{wavedef}), (\ref{E1second}) for
perturbed equations in first orders. Structure
of the general expression (\ref{gen1}) is such that {\it the deformation perturbation in any order is proportional to the corresponding
derivative of the unperturbed operator of the equation of motion.}
Since the operator has zero value on solutions, {\it all orders of deformation perturbations are identically equal to zero on the solution
of the unperturbed equation
of motion}.
Physically this means the universality of the dynamic property
covariance of scalar systems. Mathematically this means
the absolute inapplicability of the standard perturbation theory for calculating deformation
corrections to the solution of unperturbed equations for such systems.

\section{Conclusion}

So, general conclusion that can be made on the basis of all previous
examples and considerations, is that the equations with
scalar dynamic variables have a special kind
of {\it stability} with respect to arbitrary diffeomorphisms. This
stability is of purely mathematical nature and is related to our
assurance that a small change in equations should
lead to a slight change in their solutions and to a perturbation methods  based on this idea. Formally,  the
series (\ref{gen1}), of course, reveals the non-covariance
of equations (\ref{GEQ}) in the sense of its definition by
the formulas (\ref{EQ})-(\ref{EQ1}). But in fact, perturbations theory immediately discovers
the curious property: perturbations of any order vanish on the solution
to  unperturbed equation. We have called this property dynamic
covariance, since non-covariant addends to the equation in accordance with (\ref{gen1}) do not
affect its solution. The same can be expressed in another way. Consider
some solution $\Phi_0 $ to the equation (\ref{GEQ}) and denote
by $ \text{EQ}(\Phi_0) $ the family of all differential
equations to which $\Phi_0 $ is a solution\footnote{In fact, real construction of the family is very difficult problem.}.
We will call this family {\it diffeovers  over $ \Phi_0 $.}
Now we can say that {\it any coordinate diffeomorphisms
conserve  the scalar non-covariant equations considered as a members
of some diffeovers $ \text{EQ}(\Phi_0) $ over solution $\Phi_0$.}

Lets  discuss this situation from different
points of view.

\begin{enumerate}
\item In a sense, the situation in question is similar to the situation
with the so-called {\it strong} and {\it weak} conservation laws in field theory
\cite{mick}. Recall that both of them are a consequences of physical action symmetry (Noether's
theorem). The strong conservation laws as relations between field variables are satisfied no matter
are the equations of motion satisfied on these variables, while  the weak
ones are satisfied  only taking into account the equations of motion arising from
principle of the least action. An example of strong conservation laws
is the electric  charge conservation\footnote{The derivation of the law of conservation of charge from gauge invariance of action does
not use the Maxwell equations, but
post factum, it turns out that charge conservation follows from the equations.} in Maxwell's theory (a consequence of the gauge
invariance), an example of weak law ---
energy-momentum conservation (a consequence of translational invariance of the action).
In the analogy we are discussing, the general covariance of the equation is similar to the strong conservation law,
while dynamic covariance to weak ones.
\item A legitimate question arises: can it really be concluded from the property of dynamic covariance that the experimental
dependencies does not affected by any deformations  of the laboratory
setup scales? This would be very strange, since it's obvious that, say,
uniform process analyzed with using bad clocks with
variable rate, no longer look as uniform. Paradox
solved with a more accurate decomposition analysis (\ref{gen1}).
If we restrict ourselves, say, to the first order of decomposition, then it is easy
to see that the perturbed equation becomes an equation of a new type:
its order as a differential equation increases by one, and
the perturbation parameter $ \epsilon $ appears at the highest derivative.
Equations of this type are called in the theory of differential equations
{\it singularly perturbed} \cite{but}. Their solution in the form of expansions
over the parameter $\epsilon $, although it may make some sense, does not converge
uniformly to the solution of the unperturbed equation. Physical reason of this picture is presence of one or more the so-called {\it boundary layers},
wherein solution to the singular equation is very different from
unperturbed one, and outside of it --- like him. To solve singular equations, a special decomposition technique has been developed
 by A.N. Tikhonov \cite{tich}, generalizing expansion
over small parameter. It includes a standard series of perturbation theory as
part of the complete decomposition. The second part related to boundary layers is generally not analytic on $\epsilon. $

Lets  illustrate the properties of the solutions of a singularly perturbed equation
using the simplest model (\ref {E1}). In fact, the perturbed equation
(\ref {E1a}) has as a solution $ x_0 (t) $ only as a particular
solution (and very special one!). The general solution is obtained if we
suppose $ \dot x-f \neq0.$
For the case  one may separate variables in
(\ref{E1a}) and integrate it:
\begin{equation}\label{E1aa}
\frac{\displaystyle\frac{d}{dt}(\dot x-f)}{\dot
x-f}=-\frac{1}{\epsilon}\Rightarrow
\dot x(t)=f(t)+Ae^{-\int\limits_{0}^t\frac{d\xi}{\epsilon(\xi)}}\Rightarrow
\end{equation}
\[
x(t)=x_0+\int\limits_{0}^t\left(f(s)+Ae^{-\int\limits_{0}^s\frac{d\xi}{\epsilon(\xi)}}\right)\, ds,
\]
where $ A $ is the integration constant. From (\ref{E1aa}) it is clear that in our case the dependence of the solution on
$ \epsilon $ will in a certain sense always be nonanalytic, so the standard perturbation theory is absolutely inapplicable.
\item Suppose that perturbation has the form of an impulse that is described by smooth
(i.e., continuously differentiable)
nonnegative function of time with support $ \text{supp}\, \epsilon(t) = [0; T] $ and satisfies the condition $ \epsilon \ll t. $ If
 $ \epsilon (t) \sim t^\alpha + o (t), $ for smoothness it is necessary that
 $\alpha>1. $ Thus, before the pulse action, the system was described by the equation (\ref{E1}), and starting from $\tau = t = 0
 $ ---
by the  equation (\ref{E1a}), while at the time $ t = 0 $ the quantities $ x (0) = x_0 (0) $ and $ \dot x (0) = \dot x_0 (0) = f (0 ) $
should be considered as initial data. From the form of the first integral (expressions
for
 $ \dot x $) in (\ref{E1aa}) it follows that to preserve the condition $ \dot
 x (0) = f (0) $ it is necessary that the expression
 \begin{equation}\label{E1ab}
Ae^{-\int\limits_{0}^t\frac{d\xi}{\epsilon(\xi)}}
 \end{equation}
 to be vanishing at $ t = 0. $ If we put $ A = 0, $ then it will be
  vanishing identically and we return to the unperturbed
  solution, which is very special. On the other hand,
  for $ \epsilon (t) \sim t^\alpha + o (t) $ with the condition $ \alpha> 1, $ the expression (\ref {E1ab})
  in the neighborhood of $t = 0 $ will have an asymptotic form
 \begin{equation}\label{E1ac}
\sim Ae^{-C/t},\quad C>0,
 \end{equation}
that is quite compatible with initial
conditions.
\item
 Exact  calculation of perturbed
solutions leads to the equation of the form:
\begin{equation}\label{exact}
\frac{d}{dt}x(t+\epsilon(t))=f(t+\epsilon(t))(1+\dot\epsilon(t)).
\end{equation}
The equation (\ref{exact}) belongs to a class of {\it differential equations with
shifted argument} \cite {norkin}, because it connects  the values of derivative of the unknown
function with the value of the given function $\epsilon(t) $ at different times. A feature of the correct formulation of the
problem for differential equations with shifted
argument is the need to specify initial conditions not
in the form of  value of the unknown function at $t=0$, but  in the form of segment of history of the system, which
can be set arbitrarily. In addition, methods for solving equations with shifted  arguments are developed to a much lesser extent than methods
for classical differential equations. In any case, for the systems that we are interested in, that part of the known methods,
which is based on a series of perturbation theory for small shift,
does not work. We see that theoretical attempts to  take into account
the fact of scales deformations, face with serious
mathematical difficulties that far surpass
seeming "harmless" \, the nature of coordinate transformations of the form:
$t\to \tau=t+\epsilon(t).$
\item In fact, the equation (\ref{E1}) wherein  $x(\tau) $ and $ f(\tau) $
are treated as scalar quantities, has the wrong "physical-geometric
  meaning"\,: the derivative $ \dot x $ of the scalar $ x $ is 1-dimensional
  vector field, not a scalar, i.e.  when changing
  parameterization $\tau = \tau (t) $ it transformed according to the vector law:
 \begin{equation}\label{vecx}
\frac{dx}{d\tau}=\frac{dx}{dt}\frac{dt}{d\tau}.
 \end{equation}
To preserve the physical-geometric meaning of the equation (\ref{E1})
it is necessary that the  $ f $ in the right-hand side is also 1-dimensional
vector (not scalar):
 \begin{equation}\label{vecx1}
f(\tau)=f(\tau(t))\frac{dt}{d\tau}.
 \end{equation}
In this case perturbed by the deformation (\ref{var1})
equation will look as follows:
 \begin{equation}\label{vecx2}
\dot x(t)-f=-\dot\epsilon(t)\dot x(t)-\epsilon(t)(\ddot x(t)-\dot
f(t)).
 \end{equation}
Now the decomposition of perturbation theory becomes correct: for
$\Delta(t)=x(t)-x_0(t)$ after substitution into (\ref{vecx2}) we obtain the equation:
 \begin{equation}\label{vecx3}
\dot\Delta(t)=-\dot\epsilon(t) f(t)\Rightarrow
\Delta(t)=\int\limits_0^t\dot\epsilon(s)f(s)\, ds.
 \end{equation}
 Lets suppose that perturbation has step-wise form:
  \begin{equation}\label{vecx4}
\epsilon(t)=\epsilon_0(\theta(t)-\theta(t-T)),\quad \epsilon_0,T\in
\mathbb{R}_+,
 \end{equation}
where $\theta(x)$ is standard Heaviside's function. With using the well known identity of functional analysis \cite{vladimir}:
 \begin{equation}\label{vecx5}
\frac{d\theta(x)}{dx}=\delta(x),
 \end{equation}
 one immediately obtain:
 \begin{equation}\label{vecx6}
\Delta(x)=
\left\{
\begin{array}{lr}
0, &t<0;\\
\epsilon_0f(0), &0\le t<T;\\
\epsilon_0(f(0)-f(T)), &t\ge T
\end{array}
\right.
 \end{equation}
--- physically meaningful result obtained within the standard perturbation theory.
Thus, {\it using  of covariant tensor equations simultaneously eliminates the property of dynamic covariance
and restores the validity of the standard perturbation theory for calculating the deformation
corrections.} The same reasoning is  applicable to the general differential
operator (\ref{E3}), discussed in the Sec. \ref{lin3}:
the coefficients $ a_i $ form a hierarchy of geometric objects with
certain transformation laws (for $ i> 1 $ --- non-tensors), whose transformation laws must be in some sense consistent
 with  the transformation law of $a_{- 1} = - f (\tau). $
\item
Equation (\ref{vecx2}) shows that, despite the loss of dynamic covariance, perturbed equations
with correct tensor dimension still contain corrections related to this kind of covariance:
in this case they cease to be leading and total perturbation loses its singular character.
Thus, our analysis, despite the incorrect tensor dimension of the considered examples, still reveals an
important property of the perturbed equations. For equations with correct tensor dimension, it could be
called {\it latent dynamic covariance,} which does not manifest itself in such equations due to the presence of
leading nonsingular perturbations, and which could manifest itself in those systems, wherein these
non-singular leading
 perturbations for some reasons turn into zero.
 \item The dynamic covariance property is not limited to the class
 equations defining the kernels of some differential operators.
 Consider for example  the Fredholm integral equation of the 1-st kind:
 \begin{equation}\label{integ}
x(\tau)=\int\limits_{-\infty}^{+\infty}K(\tau,\tau')x(\tau')\,d\tau'\equiv \hat K[x](\tau),
 \end{equation}
where $K(\tau,t)$ is the kernel of integral operator  $\hat K,$ satisfying the condition:
\begin{equation}\label{integ1}
\int\limits_{-\infty}^{+\infty}\int\limits_{-\infty}^{+\infty}|K(\tau,\tau')|\,d\tau'\,d\tau<\infty.
\end{equation}
Under the deformation (\ref{var1}) of $\tau$ in (\ref{integ}) we obtain the decomposition:
 \begin{equation}\label{integ2}
x(t)+\epsilon\dot x(t)=\hat K[x](t)+\epsilon(t)\frac{\partial}{\partial t}\hat
K[x](t)+
\end{equation}
\[
\int\limits_{-\infty}^{+\infty}K_{,t'}(t,t')\epsilon(t')x(t')\,dt'+\hat
K[(\epsilon x)_{,t}](t).
\]
Integrating by parts the  third term in the right-hand side of (\ref{integ2})
and taking into account  $K(t,t')\stackrel{t,t'\to\pm\infty}{\rightarrow}0,$ after reducing similar terms we obtain:
 \begin{equation}\label{integ3}
x(t)-\hat K[x](t)=-\epsilon(t)\left(\dot x(t)-\frac{\partial}{\partial t}\hat
K[x](t)\right)
\end{equation}
--- perturbation proportional to  derivative of the original
integral equation, which indicates conservation
of dynamic covariance for integral equations
considered type.
\item All of the examples considered together show a somewhat unusual
symmetry type of a wide class of equations: {\it solutions to these equations possess  some kind of stability with respect to any coordinate diffeomorphisms}.
Obviously, this type of symmetry has nothing to do with Noether's
symmetries, because, firstly, not all of differential equations can be derived from the variational principle, and secondly,
those equations that are derived from the least action principle,
usually have "correct" \, transformational properties (i.e. they are covariant), which
as we saw above, as a rule eliminates the property
of dynamic covariance.
\end{enumerate}

In addition to the purely mathematical properties and consequences of the dynamic
covariance of physical equations, it must be taken into account in theory
of errors of measuring equipment, related to some aspects
of its interaction with the environment. If this interaction may be
parameterized by  dependency of the form (\ref{var1}), (\ref{deftx}) or
like them, then, as our analysis reveals, the calculation of expected
deviations sometimes cannot be taken into account by standard perturbation theory methods, even if such an interaction is small.
This circumstance is rather unexpected for experimental
physics. Note that the interaction of gravitational wave detectors
or interferometric systems with a gravitational wave field is not related to these cases,
since the perturbations of metrics of space-time that carry
gravitational waves in GR are not reduced to coordinate
diffeomorphisms.


\end{document}